\documentclass[runningheads]{llncs}
\usepackage{graphicx}
\usepackage{xcolor}
\usepackage{amsmath,amssymb,amsfonts,mathtools}
\usepackage{ stmaryrd }
\usepackage{tikz}
\usetikzlibrary{arrows,decorations.pathmorphing,backgrounds,positioning,fit,petri,knots,shapes,automata, arrows.meta, positioning}
\usepackage{standalone}

\newtheorem{innercustomprop}{Proposition}
\newenvironment{myprop}[1]
  {\renewcommand\theinnercustomprop{#1}\innercustomprop}
  {\endinnercustomprop}

\newtheorem{innercustomlem}{Lemma}
\newenvironment{mylem}[1]
  {\renewcommand\theinnercustomlem{#1}\innercustomlem}
  {\endinnercustomlem}
  
\newtheorem{innercustomcor}{Corollary}
\newenvironment{mycor}[1]
  {\renewcommand\theinnercustomcor{#1}\innercustomcor}
  {\endinnercustomcor}  
  
\newcommand{\WP}{\textnormal{WP}}
\newcommand{\isnake}{\infty\textnormal{-\textsc{snake}}}
\newcommand{\Reach}{\textsc{Reach}}
\newcommand{\ouro}{\textnormal{\textsc{ouroboros}}}

\newcommand{\Geo}{\textnormal{Geo}}
\newcommand{\MSO}{\textnormal{MSO}}

\newcommand{\N}{\mathbb{N}}
\newcommand{\om}{\omega}
\newcommand{\Z}{\mathbb{Z}}

\newcommand{\A}{\mathcal{M}}

\newcommand{\F}{\mathbb{F}}
\newcommand{\Fo}{\mathcal{F}}

\newcommand{\llangle}{\langle\langle}
\newcommand{\rrangle}{\rangle\rangle}
\newcommand{\factor}{\sqsubseteq}

\definecolor{rouge}{RGB}{255,77,77}
\definecolor{vert}{RGB}{0,178,102}
\definecolor{jaune}{RGB}{255,255,0}
\definecolor{violet}{RGB}{208,32,144}
\definecolor{orange}{RGB}{255,140,0}
\definecolor{bleu}{RGB}{0,0,205}

\newcommand{\wang}[6]{
\draw [black,fill=#3] (#1,#2+1)  -- (#1+0.5,#2+0.5) -- (#1+1,#2+1) -- cycle;
\draw [black,fill=#4] (#1+1,#2+1)  -- (#1+0.5,#2+0.5) -- (#1+1,#2) -- cycle;
\draw [black,fill=#5] (#1,#2)  -- (#1+0.5,#2+0.5) -- (#1+1,#2) -- cycle;
\draw [black,fill=#6] (#1,#2)  -- (#1+0.5,#2+0.5) -- (#1,#2+1) -- cycle;
}

\newcommand{\marker}{\vbox to 12pt{\hbox to 15pt{
\hspace{-2pt}\includegraphics[width=15pt]{figures/youRhere.png}
}}}

\begin{document}
\title{Domino Snake Problems on Groups}
%
%
\author{Nathalie Aubrun\inst{1}\orcidID{0000-0002-2701-570X} \and
Nicolas Bitar\inst{1}\orcidID{0000-0002-3460-9442}}
\authorrunning{N. Aubrun and N. Bitar}
%
\institute{
Université Paris-Saclay, CNRS, LISN, 91190 Gif-sur-Yvette, France\\
\email{\{nathalie.aubrun,nicolas.bitar\}@lisn.fr}}
\maketitle              
\begin{abstract}
 In this article we study domino snake problems on finitely generated groups. We provide general properties of these problems and introduce new tools for their study. The first is the use of symbolic dynamics to understand the set of all possible snakes. Using this approach we solve many variations of the infinite snake problem including the geodesic snake problem for certain classes of groups. Next, we introduce a notion of embedding that allows us to reduce the decidability of snake problems from one group to another. This notion enable us to establish the undecidability of the infinite snake and ouroboros problems on nilpotent groups for any generating set, given that we add a well-chosen element. Finally, we make use of monadic second order logic to prove that domino snake problems are decidable on virtually free groups for all generating sets.
 
\end{abstract}
\section{Introduction}
Since their introduction more than 60 years ago \cite{wang1961proving}, domino problems have had a long history of providing complexity lower bounds and undecidability of numerous decision problems \cite{berger1966undecidability,harel1985recurring,van2019convenience,gradel1990domino}. The input to these problems is a set of \emph{Wang tiles}: unit square tiles with colored edges and fixed orientation. The decision problems follow the same global structure; given a finite set of Wang tiles, is there an algorithm to determine if they tile a particular shape or subset of the infinite grid such that adjacent tiles share the same color along their common border? An interesting variant of this general formula are domino snake problems. First introduced by Myers in 1979 \cite{myers1979decidability}, snake problems ask for threads --or snakes-- of tiles that satisfy certain constraints. In particular, three of them stand-out. The infinite snake problem asks is there exists a tiling of a self-avoiding bi-infinite path on the grid, the ouroboros problem asks if there exists a non-trivial cycle on the grid, and the snake reachability problem asks if there exists a tiling of a self-avoiding path between two prescribed points. Adjacency rules are only required to be respected along the path. These problems have had their computability completely classified \cite{etzion1991solvability,etzion1994solvability,ebbinghaus1982undecidability,Ebbinghaus_1987,adleman2002decidability,kari2002infinite} (see Theorem \ref{thm:Z2}). In this article, we expand the scope of domino snake problems to finitely generated groups, as has been done for other domino problems \cite{aubrun2018domino}, to understand how the underlying structure affects computability. 

We present three novel ways in which to approach these problems. The first is the use of symbolic dynamics to understand the set of all possible snakes. Theorem \ref{thm:spooky} states that when this set is defined through a regular language of forbidden patterns, the infinite snake problem becomes decidable. Using this approach we solve many variations of the infinite snake problem including the geodesic snake problem for some classes of groups. Next, we introduce a notion of embedding that allows us to reduce the decidability of snake problems from one group to another. This notion enable us to establish the undecidability of the infinite snake and ouroboros problems on a large class of groups --that most notably include nilpotent groups-- for any generating set, provided that we add a central torsion-free element. Finally, to tackle virtually free groups, we express the three snake problems in the language of Monadic Second Order logic. Because for this class of groups this fraction of logic is decidable, we show that our three decision problems are decidable independently of the generating set.


\section{Preliminaries}

Given a finite alphabet $A$, we denote by $A^n$ the set of words on $A$ of length $n$, and $A^*$ the set of all finite length words including the empty word $\epsilon$. Furthermore, we denote by $A^+ = A^*\setminus\{\epsilon\}$ the set of non-empty finite words over $A$. A factor $v$ of a word $w$ is a contiguous subword; we denote this by $v \factor w$. We denote discrete intervals by $\llbracket n,m \rrbracket = \{n, n+1, ..., m-1, m\}$. We also denote the free group defined from the free generating set $S$ by $\F_S$. The proofs missing in the main text are contained in the Appendix.

\subsection{Symbolic Dynamics}
Given a finite alphabet $A$, we define the \emph{full-shift} over $A$ as the set of configurations $A^{\Z} = \{x:\Z\to A\}$. There is a natural $\Z$-action on the full-shift called the \emph{shift}, $\sigma:A^{\Z}\to A^{\Z}$, given by $\sigma(x)_{i} = x_{i+1}$. The full-shift is also endowed with the prodiscrete topology, making it a compact space.

Let $F\subseteq\Z$ be a finite subset. A \emph{pattern} of support $F$ is an element $p\in A^F$. We say a pattern $p\in A^F$ appears in a configuration $x\in A^{\Z}$, denoted $p\factor x$, if there exists $k\in\Z$ such that $x_{k+i} = p_i$ for all $i\in F$. Given a set of patterns $\Fo$, we can define the set of configurations where no pattern from $\Fo$ appears,
$$X_{\Fo} \coloneqq \{x\in A^{\Z}\mid \forall p\in\mathcal{F}, \  p \text{ does not appear in } x \}.$$

A \emph{subshift} is a subset of the full-shift $X\subseteq A^\Z$ such that there exists a set of patterns $\Fo$ that verifies $X = X_{\Fo}$. A classic result shows subshifts can be equivalently defined as closed $\sigma$-invariant subsets of the full-shift. We say a subshift $X_{\Fo}$ is
\begin{itemize}
    \item a \emph{subshift of finite type} (SFT) if $\Fo$ is finite,
    \item \emph{sofic} if $\Fo$ is a regular language,
    \item \emph{effective} if $\Fo$ is a decidable language.
\end{itemize}
Each class is strictly contained within the next. Sofic subshifts can be equivalently defined as the set of bi-infinite walks on a labeled finite graph. It is therefore decidable, given the automaton or graph that defines the subshift, to determine if the subshift is empty. Similarly, given a Turing machine for the forbidden language of an effective subshift, we have a semi-algorithm to determine if the subshift is empty.\\

Lastly, we say a configuration $x\in X$ is \emph{periodic} if there exists $k\in\Z^*$ such that $\sigma^{k}(x) = x$. We say the subshift $X$ is \emph{aperiodic} if it contains no periodic configurations.
For a comprehensive introduction to one-dimensional symbolic dynamics we refer the reader to \cite{lind2021introduction}.

\subsection{Combinatorial Group Theory}

Let $G$ be a finitely generated (f.g.) group and $S$ a finite generating set. Elements in the group are represented as words on the alphabet $S\cup S^{-1}$ through the evaluation function $w\mapsto \overline{w}$. Two words $w$ and $v$ represent the same element in $G$ when $\overline{w} = \overline{v}$, and we denote it by $w =_G v$. We say a word is \emph{reduced} if it contains no factor of the form $ss^{-1}$ or $s^{-1}s$ with $s\in S$.
 
 \begin{definition}
  Let $G$ be a group, $S$ a subset of $G$ and $R$ a language on $S\cup S^{-1}$. We say $(S, R)$ is a \emph{presentation} of $G$, denoted $G = \langle S \mid R\rangle$, if the group is isomorphic to  $\langle S \mid R\rangle = \F_S/\llangle R\rrangle$, where $\llangle R\rrangle$ is the normal closure of $R$, i.e. the smallest normal subgroup containing $R$. We say $G$ is \emph{recursively presented} if there exists a presentation $(S,R)$ such that $S$ is finite and $R$ is recursively enumerable.
 \end{definition}
For a group $G$ and a generating set $S$, we define:
  $$\WP(G, S) \coloneqq \{w\in (S\cup S^{-1})^{*} \ | \ \overline{w}=1_G\}.$$
 \begin{definition}
  The \emph{word problem} (WP) of a group $G$ with respect to a set of generators $S$ asks to determine, given a word $w\in (S\cup S^{-1})^*$, if $w\in \WP(G,S)$.
 \end{definition}
 We say a word $w\in (S\cup S^{-1})^+$ is \emph{G-reduced} if $w$ contains no factor in $\WP(G,S)$.  We say a word $w\in (S\cup S^{-1})^*$ is a \emph{geodesic} if for all words $v\in (S\cup S^{-1})^*$ such that $\overline{w} = \overline{v}$ we have $|w|\leq |v|$. For a given group $G$ and generating set $S$, we denote its language of geodesics by $\Geo(G,S)$.

We say an element $g\in G$ has \emph{torsion} if there exists $n\geq 1$ such that $g^n = 1_G$. If there is no such $n$, we say $g$ is \emph{torsion-free}. Analogously, we say $G$ is a torsion group if all of its elements have torsion. Otherwise if the only element of finite order is $1_G$ we say the group is torsion-free.

Finally, let $\mathcal{P}$ be a class of groups (for example abelian groups, free groups, etc). We say a group $G$ is \emph{virtually} $\mathcal{P}$, if there exists a finite index subgroup $H\leq G$ that is in $\mathcal{P}$.

\section{Snake Behaviour}

Although the original snake problems were posed for the Wang tile model, we make use of a different, yet equivalent, formalism (see the Appendix for a proof of the equivalence of the two models).
\begin{definition}
A tileset graph (or tileset) for a f.g. group $(G,S)$ is a finite multigraph $\Gamma = (A, B)$ such that each edge is uniquely determined by $(a,a',s)$ where $s\in S\cup S^{-1}$ and $a,a'\in A$ are its initial and final vertices respectively, and if $(a,a',s)\in B$ then $(a',a,s^{-1})\in B$.
\end{definition}
In what follows, $I$ denotes $\Z$, $\N$, or a discrete interval $\llbracket n,m\rrbracket$, depending on the context.

\begin{definition}
\label{def:snake}
Let $(G,S)$ be a f.g. group, and $\Gamma = (A,B)$ a tileset for the pair. A \emph{snake} or $\Gamma$-snake is a pair of functions $(\omega,\zeta)$, where $\omega:I\to G$ is an injective function, referred to as the snake's \emph{skeleton} and $\zeta:I\to A$ the snake's \emph{scales}. These pairs must satisfy that $d\omega_i \coloneqq \om(i)^{-1}\om(i+1)\in S\cup S^{-1}$ and $(\zeta(i), \zeta(i+1))$ must be an edge in $\Gamma$ labeled by $d\om_i$.
\end{definition}
\vspace{-0.5cm}
\begin{figure}[!ht]
    \centering
    \begin{tikzpicture}[scale=0.5,node distance = 3cm, on grid, auto]

\begin{scope}[shift={(-4,0)}]
\draw (-1,0.5) node{$t_1=$};
\draw (-1,-1.5) node{$t_2=$};

\wang{0}{0}{vert}{vert}{bleu}{jaune} 
\wang{0}{-2}{bleu}{bleu}{rouge}{vert} 
\end{scope}

\node (t1) [state] at (0,-0.5) {$t_1$};
\node (t2) [state] at (3,-0.5) {$t_2$};

\path [-stealth, thick]
    (t1) edge [bend left] node {$(1,0)$} (t2)
    (t2) edge [bend left] node {$(0,1)$} (t1);

\begin{scope}[shift={(3,0)}]
\wang{3}{-4}{vert}{vert}{bleu}{jaune} 
\wang{4}{-4}{bleu}{bleu}{rouge}{vert} 
\wang{4}{-3}{vert}{vert}{bleu}{jaune} 
\wang{5}{-3}{bleu}{bleu}{rouge}{vert} 
\wang{5}{-2}{vert}{vert}{bleu}{jaune} 
\wang{6}{-2}{bleu}{bleu}{rouge}{vert} 
\wang{6}{-1}{vert}{vert}{bleu}{jaune} 
\wang{7}{-1}{bleu}{bleu}{rouge}{vert}  
\wang{7}{0}{vert}{vert}{bleu}{jaune} 
\wang{8}{0}{bleu}{bleu}{rouge}{vert} 
\wang{8}{1}{vert}{vert}{bleu}{jaune} 
\draw (9.5,1.5) node{$\dots$};
\end{scope}

\end{tikzpicture}
    \vspace{-1cm}
    \caption{Two Wang tiles that do not tile $\Z^2$, the corresponding graph $\Gamma$ for the generating set $\left\{ (1,0),(0,1)\right\}$ (edges labeled with generator inverses are omitted for more readability) and a $\Gamma$-snake with $I=\N$. In the Wang tile model, two tiles can be placed next to each other if they have the same color on the shared side.}
    \label{fig:example_snake_Z2}
\end{figure}
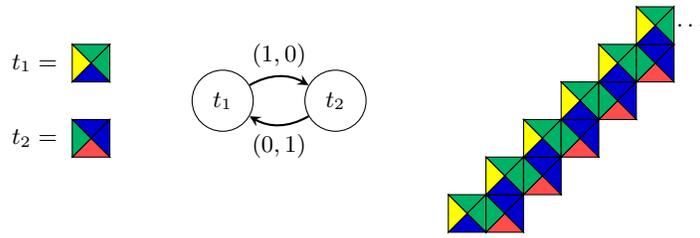
We say a snake $(\omega, \zeta)$ connects the points $p,q\in G$ if there exists a $n\in\N$ such that $(\om, \zeta)$ is defined over $\llbracket 0, n\rrbracket$, $\omega(0) = p$ and $\omega(n) = q$. We say a snake is bi-infinite if its domain is $\Z$. A $\Gamma$-ouroboros is a $\Gamma$-snake defined over $\llbracket 0,n\rrbracket$, with $n\geq 2$, that is injective except for $\om(0) = \om(n)$. In other words, a $\Gamma$-ouroboros is a well-tiled simple non-trivial cycle.
We study the following three decision problems:
\begin{definition}
Let $(G,S)$ be a f.g. group. Given a tileset $\Gamma$ for $(G,S)$ and two points $p,q\in G$, 
\begin{itemize}
    \item the \emph{infinite snake problem} asks if there exists a bi-infinite $\Gamma$-snake,
    \item the \emph{ouroboros problem} asks if there exists a $\Gamma$-ouroboros,
    \item the \emph{snake reachability problem} asks if there exists a $\Gamma$-snake connecting $p$ and $q$.
\end{itemize}
\end{definition}

We can also talk about the \emph{seeded} variants of these three problems. In these versions, we add a selected tile $a_0\in A$ to our input and ask for the corresponding snake/ouroboros to satisfy $\zeta(0) = a_0$.\\

All of these problems have been studied and classified for $\Z^2$ with its standard generating set $\left\{ (1,0),(0,1)\right\}$.
\begin{theorem}
\label{thm:Z2}
Let $S$ be the standard generating set for $\Z^2$. Then,
\begin{enumerate}
    \item The snake reachability problem for $(\Z^2, S)$ is \textrm{PSPACE}-complete \cite{etzion1994solvability},
    \item The infinite snake problem for $(\Z^2, S)$ is $\Pi^0_1$-complete \cite{adleman2002decidability},
    \item The ouroboros problem for $(\Z^2, S)$ is $\Sigma^0_1$-complete \cite{ebbinghaus1982undecidability,kari2002infinite}.
\end{enumerate}
In addition, the seeded variants of these problems are undecidable \cite{ebbinghaus1982undecidability}. 
\end{theorem}

Our aim is to extend these results to larger classes of groups and different generating sets.

\subsection{General Properties}

Let $(G,S)$ be a f.g. group and $\Gamma$ a tileset. If there exists a snake $(\omega,\zeta)$, then for every $g\in G$, $(g\om, \zeta)$ is a snake. If we define $\tilde{\om}(i) = g \om(i)$, then $d\tilde{\om} = d\om$, as the adjacency of $\zeta$ in $\Gamma$ remains unchanged. In particular, there exists a snake $(\om', \zeta)$ such that $\om'(0) = 1_G$, i.e. we may assume that a snake starts at the identity $1_G$.

The next result is a generalization of a result due to Kari for $\Z^2$. Although, the proof is essentially the same, we provide it in the Appendix for completion.
\begin{proposition}
\label{prop:compactness_for_snakes}
Let $\Gamma$ be a tileset for a f.g. group $(G,S)$. Then, the following are equivalent:
\begin{enumerate}
    \item $\Gamma$ admits a bi-infinite snake,
    \item $\Gamma$ admits a one-way infinite snake,
    \item $\Gamma$ admits a snake of every length.
\end{enumerate}
\end{proposition}

This result implies that a tileset that admits no snakes will fail to tile any snake bigger than a certain length. Therefore, if we have a procedure to test snakes of increasing length, we have a semi-algorithm to test if a tileset does not admit an infinite snake.
\begin{corollary}
\label{cor:pi01}
    If $G$ has decidable WP, the infinite snake problem is in $\Pi^0_1$.
\end{corollary}

A similar process can be done for the ouroboros problem.

\begin{proposition}
    If $G$ has decidable WP, the ouroboros problem is in $\Sigma^0_1$
    .
\end{proposition}

\begin{proof}
   Let $\Gamma$ be a tileset graph for $(G,S)$. For each $n\geq 1$, we test each word of length $n$ to see if it defines a simple loop and if it admits a valid tiling. More precisely, for $w\in(S\cup S^{-1})^{n}$, we use the word problem algorithm to check if $w$ is $G$-reduced and evaluates to $1_G$. If it is reduced, we test all possible tilings by $\Gamma$ of the path defined by following the generators in $w$. If we find a valid tiling, we accept. If not, we keep iterating with the next word length $n$ and eventually with words of length $n+1$.

   If there is a $\Gamma$-ouroboros, this process with halt and accept. Similarly, if the process halts we have found a $\Gamma$-ouroboros. Finally, if there is no $\Gamma$-ouroboros the process continues indefinitely.
\end{proof}

We also state reductions when working with subgroups or adding generators.
\begin{lemma}
\label{lem:subgroup}
Let $(G,S)$ be a f.g. group, $(H,T)$ a f.g. subgroup of $G$ and $w\in (S\cup S^{-1})^+$. Then, 
\begin{itemize}
    \item The infinite snake, ouroboros and reachability problems in $(H,T)$ many one-reduce to their respective analogues in $(G,S\cup T)$.
    \item The infinite snake, ouroboros and reachability problems in $(G,S)$ many one-reduce to their respective analogues in $(G,S\cup\{w\})$.
\end{itemize}
\end{lemma}

\begin{proof}
    Any tileset graph for $(H,T)$ is a tileset graph for $(G,S\cup T)$, and any tileset graph for $(G,S)$ is a tileset graph for $(G, S\cup\{w\})$.
\end{proof}

\section{Ossuary}

Much of the complexity of snakes comes from the paths they define on the underlying group. It stands to reason that understanding the structure of all possible injective bi-infinite paths on the group can shed light on the computability of the infinite snake problem. Let $G$ be a f.g. group with $S$ a set of generators. The \emph{skeleton} subshift of the pair $(G,S)$ is defined as 
$$ X_{G,S} \coloneqq \{x\in (S\cup S^{-1})^{\Z} \mid \forall w\sqsubseteq x, \ w\not\in\WP(G,S)\}.$$

This subshift is the set of all possible skeletons: recall from Definition \ref{def:snake} that for any skeleton $\om$, we can define $d\om:\Z\to S\cup S^{-1}$ as $d\om_i = \om(i)^{-1}\om(i+1)$. Thus, for any infinite snake $(\omega, \zeta)$: $d\omega\in X_{G,S}$. 

This formalism allows us to introduce variations of the infinite snake problem where we ask for additional properties on the skeleton. We say a subset $Y\subseteq X_{G,S}$ is \emph{skeletal} if it is shift-invariant. In particular, all subshifts of $X_{G,S}$ are skeletal.
\begin{definition}
Let $Y$ be a skeletal subset. The \emph{$Y$-snake problem} asks, given a tileset $\Gamma$, does there exist a bi-infinite $\Gamma$-snake $(\omega, \zeta)$ such that $d\om\in Y$?
\end{definition}

\subsection{Skeletons and Decidability}

A snake $(\om, \zeta)$ can be seen as two walks that follow the same labels: $\om$ is a self-avoiding walk on the Cayley graph of the group, and $\zeta$ a walk on the tileset graph. The next result, which is a direct consequence of the definitions, uses this fact to construct a single object that captures both walks.

For $(G,S)$ a f.g. group, $\Gamma$ a tileset graph, denote $X_\Gamma\subseteq\left(S\cup S^{-1}\right)^\Z$ the subshift whose configurations are the labels of bi-infinite paths over $\Gamma$. This implies $X_\Gamma$ is a sofic subshift.

\begin{proposition}
\label{thm:spooky}
    Let $(G,S)$ be a f.g. group, $\Gamma$ a tileset graph and $Y$ a non-empty skeletal subset. Then $X=Y \cap X_\Gamma$ is non-empty if and only if there is a bi-infinite $\Gamma$-snake $(\omega, \zeta)$ with $d\omega\in Y$. In addition, if $Y$ is an effective/sofic subshift, then $X$ is an effective/sofic subshift.   
\end{proposition}

The previous result reduces the problem of finding an infinite $Y$-snake, to the problem of emptiness of the intersection of two one-dimensional subshifts. As previously stated determining if a subshift is empty is co-recursively enumerable for effective subshifts, and decidable for sofics. Therefore, we can provide a semi-algorithm when the skeleton is effective. This is true for the class of recursively presented groups. Because these groups have recursively enumerable word problem, $\WP(G,S)$ is recursively enumerable for all finite generating sets. This enumeration gives us an enumeration of the forbidden patterns of our subshift.

\begin{proposition}
    Let $G$ be a recursively presented group. Then $X_{G,S}$ is effective for every finite generating set $S$.
\end{proposition}

This allows us to state the following proposition.

\begin{proposition}
\label{prop:skeletal}
Let $Y$ be a skeletal subshift. Then, if $Y$ is sofic (resp. effective) the $Y$-snake problem is decidable (resp. in $\Pi_1^0$). In particular, if $G$ is recursively presented, the infinite snake problem for any generating set is in $\Pi_1^0$. 
\end{proposition}
 
We now identify where the undecidability of the infinite snake problem comes from. If we restrict the directions in which the snake moves to 2 or 3, the problem becomes decidable. This means that the ability to make ``space-filling"-like curves in $\Z^2$ is, heuristically speaking, required for the proof of the undecidability of the infinite snake problem.

\begin{theorem}
    The infinite snake problem in $\Z^2$ restricted to 2 or 3 directions among $(1,0),(0,1),(-1,0),(0,-1)$ is decidable.
\end{theorem}

\begin{proof}
    The set of skeletons of snakes restricted to 3 directions, for instance left, right and up (denoted by $a^{-1}$, $a$ and $b$ respectively), is the subshift $Y_3 \subseteq \{a,a^{-1},b\}^{\Z}$ where the only forbidden words are $aa^{-1}$ and $a^{-1}a$. As $Y_3$ is a skeletal SFT of $X_{\Z^2, \{a,b\}}$, by Proposition \ref{prop:skeletal}, the $Y_3$-snake problem is decidable. The case of two directions is analogous as $Y_2$ is the full shift on the two generators $a$ and $b$.
\end{proof}

A natural variation of the infinite snake problem, from the point of view of group theory, is asking if there is an infinite snake whose skeleton defines a geodesic. These skeletons are captured by the geodesic skeleton subshift; a subshift of $X_{G,S}$ comprised exclusively of bi-infinite geodesic rays. Formally,
\[X^g_{G,S} = \{x\in X_{G,S} \mid \forall w\sqsubseteq x, w'=_G w : \ |w|\leq |w'|\}.\]

This subshift can be equivalently defined through the set of forbidden patterns given by $\Geo(G,S)$. Then, by the definition of a sofic shift, we can state the following proposition.

\begin{proposition}
\label{prop:geod_sofic}
    Let $(G,S)$ be a f.g. group. If $\Geo(G,S)$ is regular, then $X_{G,S}^{g}$ is sofic.
\end{proposition}

$\Geo(G,S)$ is known to be regular for all generating sets in abelian groups \cite{neumann1995automatic} and hyperbolic groups \cite{Epstein1992}, and for some generating sets in many other classes of groups \cite{neumann1995automatic,howlett1993miscellaneous,charney2004language,holt2012artin,antolin2016finite}. Theorem \ref{thm:spooky} implies that the geodesic infinite snake problem is decidable for all such $(G,S)$; most notably for $\Z^2$ with its standard generating set.

 \begin{theorem}
     The geodesic infinite snake problem is decidable for any f.g. group $(G,S)$ such that $\Geo(G,S)$ is regular. In particular, it is decidable for abelian and hyperbolic groups for all generating sets.
 \end{theorem}

What happens with skeletal subsets that are not closed and/or not effective? In these cases, Ebbinghaus showed that the problem can be undecidable outside of the arithmetical hierarchy \cite{Ebbinghaus_1987}. If we define $Y$ to be the skeletal subset of $(\Z^2, \{a,b\})$ of skeletons that are not eventually a straight line, then, $Y$ is not closed and deciding if there exists a $Y$-skeleton snake is $\Sigma^1_1$-complete. Similarly, if we take the $Y$ to be the set of non-computable skeletons of $\Z^2$, the $Y$-skeleton problem is also $\Sigma^1_1$-complete.

\section{Snake Embeddings}

Let us introduce a suitable notion of embedding, that guarantees a reduction of snake problems. To do this, we will make use of a specific class of finite-state transducer called \emph{invertible-reversible transducer}, that will translate generators of one group to another in an automatic manner.

\begin{definition}
An invertible-reversible transducer is a tuple $\A = (Q, S, T, q_0,\delta, \eta)$ where,
\begin{itemize}
    \item $Q$ is a finite set of states,
    \item $S,T$ are finite alphabets,
    \item $q_0\in Q$ is an initial state,
    \item $\delta:Q\times S\to Q$ is a transition function,
    \item $\eta:Q\times S \to T$ is such that $\eta(q,\cdot)$ is an injective function for all $q\in Q$,
\end{itemize}
such that for all $q\in Q$ and $s\in S$ there exists a unique $q'$ such that $\delta(q',s) = q$.
\end{definition}

We extend both $\eta$ and $\delta$ to manage inverses of $S$ by setting $\eta(q,s^{-1}) = \eta(q', s)^{-1}$ and $\delta(q,s^{-1}) = q'$, where $q'$ is the unique state satisfying $\delta(q',s) = q$. Furthermore, we denote by $q_w$ the state of $\A$ reached after reading the word $w\in (S\cup S^{-1})^*$ starting from $q_0$. We introduce the function $f_{\A}:(S\cup S^{-1})^*\to (T\cup T^{-1})^*$ recursively defined as $f_{\A}(\epsilon) = \epsilon$ and $f_{\A}(ws^{\pm1}) = f_{\A}(w)\eta(q_w, s^{\pm1})$.
\begin{definition}
    \label{def:snake_emb}
   Let $(G,S)$ and $(H,T)$ be two f.g. groups. A map $\phi:G\to H$ is called a \emph{snake embedding} if there exists a transducer $\A$ such that $\phi(g) = f_{\A}(w)$ for all $w\in (S\cup S^{-1})^*$ such that $\bar{w} = g$, and $f_{\A}(w) =_H f_{\A}(w')$ if and only if $w=_G w'$.
\end{definition}

\begin{remark}
Snake-embeddings are a strictly stronger form of a translation-like action. Such an action is a right group action $*:G\to H$ that is free, i.e. $h*g = h$ implies $g = 1_G$, and $\{d(h,h\ast g)\mid h\in H\}$ is bounded for all $g\in G$. A straightforward argument shows that if $\phi:(G,S)\to (H,T)$ is a snake-embedding, then $h\ast g = h\phi(g)$ is a translation-like action. The converse is not true: there are translation-like actions that are not defined by snake-embeddings. For instance, from Definition \ref{def:snake_emb} we see that there is a snake embedding from $\Z$ to a group $G$ if and only if $\Z$ is a subgroup of $G$. Nevertheless, infinite torsion groups admit translation-like actions from $\Z$, as shown by Seward in \cite{Seward_2014}, but do not contain $\Z$ as  a subgroup.
\end{remark}


\begin{proposition}
\label{prop:embedding}
    Let $(G,S)$ and $(H,T)$ be two f.g. groups such that there exists a snake-embedding $\phi:G\to H$. Then, the infinite snake (resp. ouroboros) problem on $(G,S)$ many-one reduces to the infinite snake (resp. ouroboros) problem on $(H,T)$.
\end{proposition}
The reduction consists in taking a tileset for $(G,S)$ and using the transducer to create a tileset for $(H,T)$ that is consistent with the structure of $G$. Because the transducer is invertible-reversible, we have a computable way to transform a bi-infinite snake from one group to the other.\\

Using snake-embeddings we can prove that non-$\Z$ f.g. free abelian groups have undecidable snake problems.

\begin{proposition}
    The infinite snake and ouroboros problems on $\Z^d$ with $d\geq 2$ are undecidable for all generating sets.
\end{proposition}

\begin{proof}
    Let $S = \{v_1, ..., v_n\}$ be a generating set for $\Z^d$. As $S$ generates the group, there are two generators $v_{i_1}$ and $v_{i_2}$, such that $v_{i_1}\Z\cap v_{i_2}\Z = \{1_{\Z^d}\}$. Then, $H = \langle v_{i_1}, v_{i_2}\rangle\simeq\Z^2$ and there is clearly a snake-embedding from $\Z^2$ to $H$. Finally, by Lemma \ref{lem:subgroup}, the infinite snake and ouroboros problems are undecidable for $(\Z^d, S)$.
\end{proof}

\section{Virtually Nilpotent Groups}

Through the use of snake-embeddings and skeleton subshifts, we extend undecidability results from abelian groups to the strictly larger class of virtually nilpotent groups. For any group $G$ we define $Z_0(G) = \{1_G\}$ and 
$$Z_{i+1}(G) = \{g\in G \mid ghg^{-1}h^{-1}\in Z_i(G), \forall h\in G\}.$$
The set $Z_1(G)$ is called the \emph{center} of $G$, and by definition is the set of elements that commute with every element in $G$. We say a group is \emph{nilpotent} if there exists $i\geq 0$ such that $Z_{i}(G) = G$.

The next Lemma is stated for a larger class of groups that contain nilpotent groups, that will allow us to prove the undecidability results on the latter class.
\begin{lemma}
\label{lem:snake_embedding_torsion}
Let $(G,S)$ be a f.g. group that contains an infinite order element $g$ in its center, such that $G/\langle g\rangle$ is not a torsion group. Then, there is a snake embedding from $(\Z^2, \{a,b\})$ into $(G,S\cup\{g\})$, where $\{a,b\}$ is the standard generating set for $\Z^2$.
\end{lemma}

The proof consists in finding a distorted copy of $\Z^2$ within $(G, S\cup\{g\})$. One of the copies of $\Z$ is given by $\langle g \rangle \simeq\Z$. The other is obtained through the following result.
\begin{proposition}
\label{prop:aperiodic}
Let $(G,S)$ be a f.g. group. Then, $G$ is a torsion group if and only if $X_{G,S}$ is aperiodic.
\end{proposition}

Using $g$ and a periodic point from this proposition we construct the snake-embedding.
 
\begin{proposition}
\label{prop:pre_nilpotent}
Let $(G,S)$ be a f.g. group that contains a infinite order element $g$ in its center and $G/\langle g\rangle$ is not a torsion group. Then, $(G,S\cup\{g\})$ has undecidable infinite snake and ouroboros problems.
\end{proposition}

\begin{proof}
    By Lemma \ref{lem:snake_embedding_torsion}, there is a snake-embedding from $\Z^2$ to $(G,S\cup\{g\})$. Combining Proposition \ref{prop:embedding} and Theorem \ref{thm:Z2}, we conclude that both problems are undecidable on $(G,S\cup\{g\})$.
\end{proof}

\begin{theorem}
Let $(G,S)$ be a f.g. infinite, non-virtually $\Z$, nilpotent group. Then there exists $g$ such that $(G,S\cup\{g\})$ has undecidable infinite snake and ouroboros problems.
\end{theorem}

\begin{proof}
    Let $G$ be a f.g. infinite nilpotent group that is not virtually cyclic. Because $G$ is nilpotent, there exists a torsion-free element $g\in Z_1(G)$. Furthermore no quotient of $G$ is an infinite torsion group \cite{Clement2017}. In addition, as $G$ is not virtually cyclic $G/\langle g\rangle$ is not finite. Therefore, by Proposition \ref{prop:pre_nilpotent}, both problems are undecidable on $(G,S\cup\{g\})$.
\end{proof}

Through Lemma \ref{lem:subgroup} we obtain undecidability for virtually nilpotent groups.
\begin{corollary}
Let $G$ be a f.g. infinite, non virtually $\Z$, virtually nilpotent group. Then there exists a finite generating set $S$ such that $(G,S)$ has undecidable infinite snake and ouroboros problems.
\end{corollary}

\section{Snakes and Logic}
 We want to express snake problems as a formula that can be shown to be satisfied for a large class of Cayley graphs. To do this we use Monadic Second-Order (MSO) logic, as has been previously been done for the domino problem. Our formalism is inspired by \cite{bartholdi2022monadic}. Let $\Lambda = (V, E)$ be an $S$-labeled graph with root $v_0$. This fraction of logic consists of variables $P,Q,R,...$ that represent subsets of vertices of $\Lambda$, along with the constant set $\{v_0\}$; as well as an operation for each $s\in S$, $P\cdot s$, representing all vertices reached when traversing an edge labeled by $s$ from a vertex in $P$. In addition, we can use the relation $\subseteq$, Boolean operators $\wedge, \vee, \neg$ and quantifiers $\forall, \exists$. For instance, we can express set equality by the formula $(P = Q)\equiv (P\subseteq Q \wedge Q\subseteq P)$ and emptiness by $(P = \varnothing) \equiv \forall Q(P\subseteq Q)$. 
We can also manipulate individual vertices, as being a singleton is expressed by $(|P| = 1) \equiv P\neq\varnothing \wedge \forall Q\subseteq P(Q = \varnothing \vee P = Q)$. For example, $\forall v\in P$ is shorthand notation for the expression $\forall Q (Q\subseteq P \wedge |Q|=1)$. Notably, we can express non-connectivity of a subset $P\subseteq V$ by the formula $\textsc{nc}(P)$ defined as
$$\exists Q\subseteq P, \exists v,v'\in P\left(v\in Q \wedge v'\not\in Q \wedge \forall u,w\in P\left(u\in Q \wedge \bigvee_{s\in S} u\cdot s = w \implies w\in Q \right)\right).$$
The set of formulas without free variables obtained with these operations is denoted by $\MSO(\Lambda)$. We say $\Lambda$ has \emph{decidable} MSO logic, if the problem of determining if given a formula in $\MSO(\Lambda)$ is satisfied is decidable.

The particular instance we are interested in is when $\Lambda$ is the Cayley graph of a f.g. group $G$ labeled by $S$ a symmetric finite set of generators, that is, $S = S^{-1}$ (we take such a set to avoid cumbersome notation in this section). In this case, the root of our graph is the identity $v_0 = 1_G$. A landmark result in the connection between MSO logic and tiling problems comes from Muller and Schupp \cite{muller1983groups,muller1985theory}, as well as Kuskey and Lohrey \cite{kuske2005logical}, who showed that virtually free groups have decidable MSO logic. Because the Domino Problem can be expressed in MSO, it is decidable on virtually free groups. Our goal is to obtain an analogous result for domino snake problems.
To express infinite paths and loops, given a tileset graph $\Gamma = (A,B)$, we will partition a subset $P\subseteq V$ into subsets indexed by $S$ and $A$, such that $P_{s,a}$ will contain all vertices with the tile $a$ that point through $s$ to the continuation of the snake. 
We denote the disjoint union as $P = \coprod_{s\in S, a\in A}P_{s, a}$. First, we express the property of always having a successor within $P$ as
$$N(P, \{P_{s,a}\}) \equiv \bigwedge_{s\in S, a\in A}\left(P_{s,a}\cdot s\subseteq P\right).$$
We also want for this path to not contain any loops, by asking for a unique predecessor for each vertex:
\begin{align*}
    \textnormal{up}(v) &\equiv \exists! s\in S,a\in A: v\in P_{s,a}\cdot s,\\
    &\equiv \left(\bigvee_{\substack{s\in S\\a\in A}} v\in P_{s,a}\cdot s\right)\wedge\left(\bigwedge_{\substack{a,a'\in A\\ a\neq a'}}\bigwedge_{\substack{s,t\in S\\ s\neq t}}\neg((v\in P_{s,a}\cdot s)\wedge (v\in P_{t,a'}\cdot t))\right).
\end{align*} Then, for a one-way infinite path
$$\textnormal{UP}(P, \{P_{s,a}\}) \equiv \forall v\in P\left((v = v_0 \wedge \bigwedge_{s\in S, a\in A}v\not\in P_{s,a}\cdot s)  \vee (v\neq v_0 \wedge \textnormal{up}(v))\right),$$
Thus, we state the property of having an infinite path as follows:
$$\infty\textsc{ray}(P,\{P_{s,a}\}) \equiv \left(v_0\in P \wedge P = \coprod_{s\in S,a\in A}P_{s,a} \wedge N(P, \{P_{s,a}\}) \wedge UP(P, \{P_{s,a}\})\right).$$
In fact, we can do a similar procedure to express the property of having a simple loop within $P$ by slightly changing the previous expressions. The only caveat comes when working with $S$ a symmetric finite generating set of some group, as we must avoid trivial loops such as $ss^{-1}$.
$$\ell(P, \{P_{s,a}\}) \equiv \forall v\in P,\textnormal{up}(v) \wedge \bigwedge_{s\in S,a,a'\in A} \left(P_{s,a}\cdot s\cap P_{s^{-1},a'} = \varnothing \right).$$
This way, admitting a simple loop is expressed as
\begin{align*}
    \textsc{loop}(P,\{P_{s,a}\}) \equiv&\left(v_0\in P \wedge P = \coprod_{s\in S,a\in A}P_{s,a} \wedge N(P, \{P_{s,a}\}) \wedge \ell(P, \{P_{s,a}\})\right) \\
    &\wedge \forall Q\subseteq P, \forall\{Q_{s,a}\}\left(\neg \infty\textsc{ray}(Q,\{Q_{s,a}\})\right).
\end{align*}

\begin{lemma}
\label{lem: mso_ray}
Let $P\subseteq V$. Then, 
\begin{enumerate}
    \item If there exists a partition $\{P_{s,a}\}_{s\in S, a\in A}$ such that $\infty\textsc{ray}(P,\{P_{s,a}\})$ is satisfied, $P$ contains an infinite injective path. Conversely, if $P$ is the support of an injective infinite path rooted at $v_0$, there exists a partition $\{P_{s,a}\}_{s\in S, a\in A}$ such that $\infty\textsc{ray}(P,\{P_{s,a}\})$ is satisfied.
    \item If there exists a partition $\{P_{s,a}\}_{s\in S, a\in A}$ such that $\textsc{loop}(P,\{P_{s,a}\})$ is satisfied, $P$ contains a simple loop. Conversely, if $P$ is the support of a simple loop based at $v_0$, there exists a partition $\{P_{s,a}\}_{s\in S, a\in A}$ such that $\textsc{loop}(P, \{P_{s,a}\})$ is satisfied. 
\end{enumerate}
\end{lemma}

With these two structure-detecting formulas, we can simply add the additional constraint that $P$ partitions in a way compatible with the input tileset graph of the problem, in the direction of the snake. This is captured by the formula
$$D_{\Gamma}(\{P_{s,a}\}) \equiv \bigwedge_{(a,a',s)\not\in B}\bigwedge_{s'\in S}P_{a,s}\cdot s \cap P_{a',s'} = \varnothing.$$
\begin{theorem}
Let $\Lambda$ be a Cayley graph of generating set $S$. The infinite snake problem, the reachability problem and the ouroboros problem can be expressed in $\MSO(\Lambda)$.
\end{theorem}

\begin{proof}
Let $\Gamma = (A, B)$ be a tileset graph for $\Lambda$. By Lemma \ref{lem: mso_ray}, it is clear that
$$\isnake(\Gamma) \equiv \exists P \exists \{P_{s,a}\}\left(\infty\textsc{ray}(P,\{P_{s,a}\}) \wedge D_{\Gamma}(\{P_{s,a}\})\right),$$
$$\ouro(\Gamma) \equiv \exists P \exists \{P_{s,a}\}\left(\textsc{loop}(P,\{P_{s,a}\}) \wedge D_{\Gamma}(\{P_{s,a}\})\right),$$
exactly capture the properties of admitting a one-way infinite $\Gamma$-snake and $\Gamma$-ouroboros respectively. Remember that Proposition \ref{prop:compactness_for_snakes} tells us that admitting a one-way infinite snake is equivalent to admitting a bi-infinite snake. 
We finish by noting that for reachability in a Cayley graph, we can take $p = v_0$. Then, verifying the formula $\Reach(\Gamma, q)$ defined as
$$\exists P \exists \{P_{s,a}\}\left(q\in P \wedge \neg\textsc{nc}(P) \wedge P = \coprod_{s\in S, a\in A}P_{s,a} \wedge UP(P,\{P_{s,a}\}) \wedge D_{\Gamma}(\{P_{s,a}\})\right),$$
is equivalent to $P$ containing the support of a $\Gamma$-snake that connects $p$ to $q$.
\end{proof}

As previously mentioned, virtually free groups have decidable MSO logic for all generating sets. Thus, we can state the following corollary.

\begin{corollary}
    Both the normal and seeded versions of the infinite snake, reachability and ouroboros problems are decidable on virtually free groups, independently of the generating set.
\end{corollary}

\begin{proof}
    Let $\Gamma = (A,B)$ be a tileset graph with $a_0\in A$ the targeted tile. Then adding the clause $\bigvee_{s\in S}v_0\in P_{s,a_0}$ to the formulas of any of the problems in question, we obtain a formula that expresses its corresponding seeded version.
\end{proof}


\section*{Acknowledgments}

We would like to thank Pierre Guillon and Guillaume Theyssier for helping with Lemma \ref{lem: mso_ray}.  We would also like to thank David Harel and Yael Etzion for providing a copy of \cite{etzion1991solvability}. We are grateful to the anonymous referees for their useful remarks.

%
%
%
\bibliographystyle{splncs04}
\bibliography{biblio}

\newpage

\section*{Appendix A: Proofs}

\begin{myprop}{\ref{prop:compactness_for_snakes}}
Let $\Gamma$ be a tileset for a f.g. group $(G,S)$. Then, the following are equivalent:
\begin{enumerate}
    \item $\Gamma$ admits a bi-infinite snake,
    \item $\Gamma$ admits a one-way infinite snake,
    \item $\Gamma$ admits a snake of every length.
\end{enumerate}
\end{myprop}

\begin{proof}
    Notice that a bi-infinite snake always contains a one-way infinite snake, and a one-way infinite snake contains snakes of arbitrary length. Therefore, it remains to prove (3)$\implies$(1). Let $(\om_n, \zeta_n)_{n\in\N}$ be a sequence of snakes with $\om_{n}:\llbracket-n,n\rrbracket\to G$ which we can take to satisfy $\om_n(0) = 1_{G}$ for all $n\in\N$. As we have an infinite amount of snakes, and for every $m\geq 1$, $B_{G}(1_G, m)$ the ball of radius $m$ is finite, we can extract a subsequence $\varphi:\N\to\N$ such that $\om_{\varphi(n)}(\llbracket -m, m\rrbracket)$ and $\zeta_{\varphi(n)}(\llbracket -m, m\rrbracket)$ coincide for all $n\in\N$. By iterating this process we obtain a bi-infinite snake $(\om,\zeta)$. 
\end{proof}

\begin{mycor}{\ref{cor:pi01}}
    If $G$ has decidable word problem, the infinite snake problem is in $\Pi^0_1$.
\end{mycor}

\begin{proof}
    Let $\Gamma=\left(V_\Gamma,E_\Gamma\right)$ be a tileset graph for $(G,S)$. We will create a recursive process that will test larger and larger snakes. Define $\mathfrak{F}^{(0)} = \{\om_0\}$ where $\om_0: \{0\} \to \{\epsilon\}$ and $\mathfrak{S}$ be the set of snakes $(\om_0, \zeta)$ with $\zeta:\{0\}\to V_{\Gamma}$. Our recursive procedure will take skeletons in $\mathfrak{F}^{(n)}$ and try to tile them without mismatches. All snakes that we obtain with this procedure will define the set $\mathfrak{S}^{(n)}$.  We proceed as follows: 
    \begin{itemize}
        \item For every $\om\in\mathfrak{F}^{(n)}$, we create functions $\om_{s,t}:\llbracket-n-1,n+1\rrbracket\to(S\cup S^{-1})^{2n+1}$ for every $s,t\in S\cup S^{-1}$, by
        $$\om_{s,t}(i) = \begin{cases}\om(i) \ \text{ if } i\in\llbracket-n, n\rrbracket \\ \om(-n)s \text{ if } i = -n-1 \\ \om(n)t \text{ if } i = n+1\end{cases}.$$
        By using the algorithm for the word problem, we select those functions such that $\om_{s,t}(\pm(n+1))$ do not create new factors that evaluate to the identity, and add them to $\mathfrak{F}^{(n+1)}$.
        \item Next, for every $\om\in\mathfrak{F}^{(n+1)}$ we test all possible tilings of $\om(\llbracket-n, n\rrbracket)$ and add the pairs that are correctly tiled to $\mathfrak{S}^{(n+1)}$.
    \end{itemize}
    Thus, we can enumerate the sets of finite snakes $\mathfrak{S}^{(n)}$. We can conclude that the infinite snake problem is $\Pi^0_1$ as Proposition \ref{prop:compactness_for_snakes} tells us that there is no bi-infinite $\Gamma$-snake if and only if there exists $n\in\N$ such that $\mathfrak{S}^{(n)} = \varnothing$.
\end{proof}

\begin{myprop}{\ref{thm:spooky}}
    Let $(G,S)$ be a f.g. group, $\Gamma$ a tileset graph and $Y$ a non-empty skeletal subset. Then $X=Y \cap X_\Gamma$ is non-empty if and only if there is a bi-infinite $\Gamma$-snake $(\omega, \zeta)$ with $d\omega\in Y$. In addition, if $Y$ is an effective/sofic subshift, then $X$ is an effective/sofic subshift.   
\end{myprop}

\begin{proof}
Let $\Gamma = (A, B)$ be a tileset for $G$, with generating set $S$. Remind that $X_\Gamma\subseteq(S\cup S^{-1})^{\Z}$ is the subshift whose configurations are the labels of bi-infinite paths over $\Gamma$. This implies $X_{\Gamma}$ is sofic subshift. Next, let $X$ be the intersection $X = Y\cap X_\Gamma$. Because $X_\Gamma$ is sofic, $X$ will be an effective (resp. sofic) subshift when $Y$ is an effective (resp. sofic) subshift.

Assume we have a bi-infinite snake $(\omega,\zeta)$ such that $d\omega \in Y$. As the pair is a $\Gamma$-snake, for all $i\in\Z$ the transition from $\zeta(i)$ to $\zeta(i+1)$ is an edge on $\Gamma$ labeled by $d\omega_i$. Therefore $d\omega\in X_\Gamma$, and as a consequence $d\omega\in X$. Conversely, suppose there exists $x\in X$. Let $\mathfrak{i}:\Z \to A$ be the function that gives us the initial vertex of every traversed edge. That is, $\mathfrak{i}(i)$ is the departure vertex for the edge traversed in $x_i$. Then, define the snake $(\om_x, \zeta_x)$ with scales $\zeta_x(i) = \mathfrak{i}(i)$, and skeleton $\omega_x:\Z\to G$ by $\omega_{x}(i) = \overline{x_{[0,i]}}$ when $ i \geq 0$ and $(\overline{x_{[i,0]}})^{-1}$ when $i < 0$. This skeleton satisfies $d\omega_x(i) = x_i$ and therefore $d\om_x\in Y$. Furthermore, because $Y\subseteq X_{G,S}$ we have that $w_x(\Z) \subseteq G$ is injective. 
\end{proof}

\begin{myprop}{\ref{prop:embedding}}
    Let $(G,S)$ and $(H,T)$ be two f.g. groups such that there exists a snake-embedding $\phi:G\to H$. Then, the infinite snake (resp. ouroboros) problem on $(G,S)$ many-one reduces to the infinite snake (resp. ouroboros) problem on $(H,T)$.
\end{myprop}

\begin{proof}
    Let $\Gamma = (A,B)$ be a tileset for $(G,S)$. We will define $\tilde{\Gamma}$, a tileset for $(H,T)$ by using the transducer $\A$ given by the snake embedding. The set of vertices is given by $\tilde{A} = A \times Q$ and we add an edge from $(u, q_1)$ to $(v, q_2)$ labeled by $t$ if and only if there is an edge $(u,v)$ in $B$ labeled by $s$, in addition to $\delta(q_1, s) = q_2$ and $\eta(q_1, s) = t$. Because $B$ is finite and $\A$ is a finite automaton, the reduction is computable.

    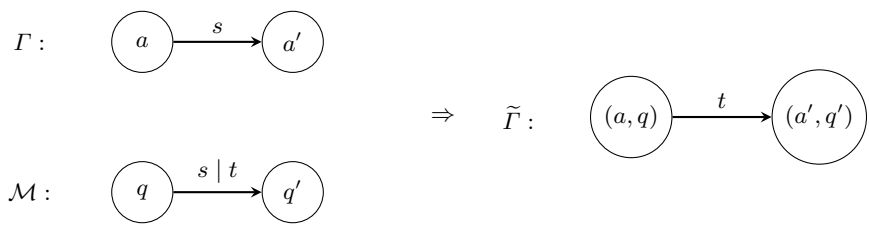
\begin{figure}[!ht]
    \centering
     
\begin{tikzpicture}[node distance = 3cm, on grid, auto]
    \node at (-1.5,0) {$\Gamma:$};
    \node (q0) [state] at (0,0) {$a$};
    \node (q1) [state] at (2,0) {$a'$};

    \node at (-1.5,-2) {$\mathcal{M}:$};
    \node (q2) [state] at (0,-2) {$q$};
    \node (q3) [state] at (2,-2) {$q'$};

    \node at (4,-1) {$\Rightarrow$};

    \node at (5,-1) {$\widetilde{\Gamma}:$};
    \node (q4) [state] at (6.5,-1) {$(a,q)$};
    \node (q5) [state] at (9,-1) {$(a',q')$};

\path [-stealth, thick]
    (q0) edge node {$s$}   (q1)
    (q2) edge node {$s\mid t$}   (q3)
    (q4) edge node {$t$}   (q5);

\end{tikzpicture}
    \caption{Reducing snake problems using a snake embedding.}
    \label{fig:reduction_snake_embedding}
    \end{figure}

    Now, let $(\omega, \zeta)$ be a $\Gamma$-snake. We define $(\tilde{\om},\tilde{\zeta})$ on $(H,T)$ by $\tilde{\om}(i) = \phi(\om(i))$ and $\tilde{\zeta}(i) = (\zeta(i), q_{\om(i)})$. As $\om$ is injective and $\phi$ is injective, $\tilde{\om}$ is injective. Furthermore, by definition there is an edge in $\tilde{\Gamma}$ from $\tilde{\zeta}(i)$ to $\tilde{\zeta}(i+1)$, as there is one from $\zeta(i)$ to $\zeta(i+1)$ in $\Gamma$, $\delta(q_{\om(i)}, d\om(i)) = q_{\om(i+1)}$ and $\eta(q_{\om(i)},d\om(i)) = d\tilde{\om}(i)$. Thus, $(\tilde{\om},\tilde{\zeta})$ is a $\tilde{\Gamma}$-snake.

    Conversely, let $(\tilde{\om},\tilde{\zeta})$ be a $\tilde{\Gamma}$-snake such that $\tilde{\om}(0) = 1_H$. Our objective is to find a $\Gamma$-snake $(\om,\zeta)$ such that $\tilde{\om} = \phi(\om)$. To do this, we introduce some notation. Let us denote $\tilde{\zeta}(i) = (q(i), u_i)$ and for any $q\in Q$ we denote by $\theta_{q}:\eta(q,S)\to S$ the inverse of the function $\eta(q,\cdot):S\to \eta(q,S)$. As $q(0)$ may not necessarily be $q_0$, let $w\in (S\cup S^{-1})^*$ such that $q_w = q(0)$ and $g$ the element $w$ represents in $G$. Without loss of generality we can change our snake $(\tilde{\om}, \tilde{\zeta})$ so that $\tilde{\om}(0) = \phi(g)$. Now, define $\om$ recursively by $\om(0) = g$ and $\om(i+1) = \om(i)\cdot\theta_{q(i+1)}(d\tilde\om(i))$.\\

    \textbf{Claim: } $q(i) = q_{w d\om(0) ... d\om(i-1)}$ and $\tilde{\om}(i) = \phi(\om(i))$.\\

    We will prove the claim by induction on $i\geq 0$, as the case for $i<0$ works analogously. The base case is clear by definition, as we imposed that $q_{w} = q(0)$ and $\phi(\om(0)) = \tilde{\om}(0)$. Next, assume our hypothesis is true for all integers up to $i$. Because $\tilde{\zeta}(i)$ is placed next to $\tilde{\zeta}(i+1)$ along the generator $d\tilde{\om}(i)\in T$, and we defined $d\om(i)$ to be $\theta_{q(i+1)}(d\tilde\om(i))$, the transition function $\delta$ sends $q(i)$ to $q(i+1)$ when reading $d\om(i)$. By the induction hypothesis $q(i)$ is the state at which we arrive after reading the word $wd\om(0) \ ... \ d\om(i-1)$, and therefore $q(i+1)$ is the state at which we arrive after reading $wd\om(0) \ ... \ d\om(i-1)d\om(i)$. Finally, we have
    \begin{align*}
        \phi(\om(i+1)) &= \phi(\om(i)\cdot d\om(i)) \\
                        &= \phi(d\om(0)\cdot ... \cdot d\om(i-1)\cdot d\om(i))\\
                       &= f_{\A}(d\om(0)\cdot ... \cdot d\om(i-1)\cdot d\om(i)) \\
                       &= \phi(\om(i))\eta(q(i+1), d\om(i)) \\
                       &= \tilde{\om}(i)\eta(q(i+1), d\om(i)).
    \end{align*}
    As we chose $d\omega(i) = \theta_{q(i+1)}(d\tilde\om(i))$, we precisely have $\eta(q(i+1), d\om(i)) = d\tilde{\om}(i)$. Thus, 
    $$\phi(\om(i+1)) = \tilde{\om}(i)d\tilde{\om}(i) = \tilde{\om}(i+1).$$

    The Claim shows that $\om$ is injective, as $\phi$ is injective. Finally, we set $\zeta(i) = u_i$, that is, the second element of the ordered pair $\tilde{\zeta}(i)$. Consequently, $(\om, \zeta)$ is a $\Gamma$-snake.

    For the ouroboros problem the same procedure applies because if $(\om,\zeta)$ is an ouroboros, then $(g\om,\zeta)$ is also an ouroboros for every $g\in G$.
\end{proof}

\begin{myprop}{\ref{prop:aperiodic}}
Let $(G,S)$ be a f.g. group. Then, $G$ is a torsion group if and only if $X_{G,S}$ is aperiodic.
\end{myprop}

\begin{proof}
    Let $g\in G$ be the torsion-free element with the smallest word length. Let $w$ be a geodesic representing $g$. Notice $w$ is cyclically reduced, if not, the cyclic reduction of $w$ would be a shorter torsion-free element. Let us prove that $w^n$ is $G$-reduced by induction over $n\geq 2$. 
    
    For the base case, suppose there exists a strict factor $w'\factor w^2$ such that $w' =_G 1_G$. Because $w$ is a geodesic, it does not contain factors that evaluate to the identity.  Therefore, $w' = uv$ with $w = w_u u = vw_v$ for two words $w_{u},w_{v}\in (S\cup S^{-1})^*$. Suppose $u$ and $v$ have different lengths, for instance $|u| < |v|$. Because $u =_{G} v^{-1}$, we have $w =_{G} w_u v^{-1}$ and $|w| > |w_u v^{-1}|$ which contradicts the fact that $w$ is a geodesic. Thus $|u| = |v|$. If their lengths are strictly bigger than $\frac{1}{2}|w|$, then the word obtained by deleting $w'$ from $w^2$ will represent a torsion-free element of length strictly smaller than $|w|$. Therefore $|u| = |v| \leq \frac{1}{2}|w|$. Then, $w$ can be written as $w = urv$ for some $r\in (S\cup S^{-1})^{*}$. But, because $v =_G u^{-1}$ we have $w =_{G} uru^{-1}$ which is a contradiction. This means $w^2$ is $G$-reduced.

    Next, assume $w^n$ is $G$-reduced for $n>2$. Suppose there is a strict factor $w' = uw^{n-1}v\factor w^{n+1}$ with $u,v\in (S\cup S^{-1})^*$, such that $w' =_G 1_G$. Because $w'$ is a strict factor, either $|u| < |w|$ or $|v| < |w|$. Without loss of generality we assume the former. We tackle two cases separately:
    \begin{itemize}
        \item If $v = w$, we have $w' = uw^{n}$. Then $u =_{G} w^{-n}$ is torsion-free and $|u|<|w|$, which is a contradiction.
        \item If $|v| < |w|$, let us write $w = w_u u = vw_v$ for two words $w_{u},w_{v}\in (S\cup S^{-1})^+$. Then, $w'$ can be written as $w' = uv(w_v v)^{n-2}$. Because $w = vw_v$ is torsion-free, $w_vv$ also is, and consequently $uv =_{G} (w_v v)^{-(n-2)}$ is torsion-free. Notice that $uv$ is $G$-reduced as it is a factor of $w^2$. Because $w$ is the smallest-torsion free element, $|u| + |v| = |uv| \geq |w|$. Similarly, as $w' =_G 1_G$ we know $w_uw_v =_G w^{n+1}$ is torsion-free and $G$-reduced (also a factor of $w^2$). Yet, $|w^2| = |w_u|+|u| + |v| + |w_v|$ which means $|w_uw_v| \leq |w|$. As $w$ is the smallest torsion-free element $|w| = |w_u| + |w_v| = |u| + |v|$. By using the fact that $w = w_u u = vw_v$ we inevitably have $w = uv$. This implies $w' = uw^{n-1}v = w^n$, which is a contradiction.
    \end{itemize}
    As $w^n$ is always $G$-reduced, the configuration $w^{\infty}$ contains no factors that evaluate to the identity and is therefore in $X_{G,S}$.\\

    Suppose $G$ is a torsion group and let $x\in X_{G,S}$ be a periodic configuration that infinitely repeats the word $w$. Let $g = \overline{w}$. By definition of the skeleton subshift, $g^n = \overline{w^n} \neq 1_G$ for all $n\in\N$. This contradicts the fact that $G$ is a torsion group.
\end{proof}

\begin{mylem}{\ref{lem:snake_embedding_torsion}}
Let $(G,S)$ be a f.g. group that contains an infinite order element $g$ in its center, such that $G/\langle g\rangle$ is not a torsion group. Then, there is a snake embedding from $\Z^2$ into $(G,S\cup\{g\})$.
\end{mylem}
\begin{proof}
    Let us take $G$ and $g$ as in the statement and $\Z^2 = \langle a,b \mid [a,b]\rangle$. Because $G/\langle g\rangle = \langle S\rangle$ is not a torsion group, by Proposition \ref{prop:aperiodic} there exists $w\in (S\cup S^{-1})^*$ such that $w^{\infty}\in X_{G/\langle g\rangle,S}$. In other words, no factors of the infinite configuration $w^{\infty}$ evaluates to $g^{n}$ for some $n\in\Z$.  We construct the snake-embedding by defining the invertible-reversible Mealy automaton $\A$ as follows. The set of states is $Q = \{q_0, ..., q_{m-1}\}$, where $m=|w|$, with transition function $\delta$ such that $\delta(q_i, a) = q_i$ and $\delta(q_i, b) = q_{(i+1\text{ mod }m)}$. The transducer is given by $\eta(q_i,a) = g$ and $\eta(q_i,b) = w_i$. These definitions guarantee that $\A$ is a transducer. 

    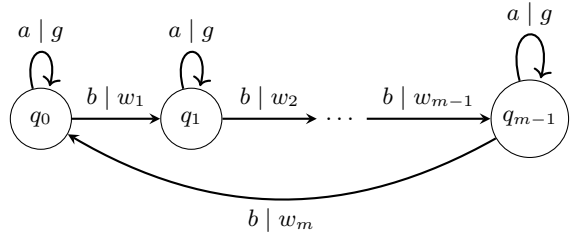
\begin{figure}[!ht]
        \centering
         
\begin{tikzpicture}[node distance = 3cm, on grid, auto]
    \node (q0) [state] at (0,0) {$q_0$};
    \node (q1) [state] at (2,0) {$q_1$};
    \node (q2) at (4,0) {$\dots$};
    \node (qn) [state] at (6.5,0) {$q_{m-1}$};

\path [-stealth, thick]
    (q0) edge node {$b\mid w_1$}   (q1)
    (q1) edge node {$b\mid w_2$}   (q2)
    (q2) edge node {$b \mid w_{m-1}$}   (qn)
    (qn) edge [bend left] node {$b \mid w_m$}   (q0)
    (q0) edge [loop above]  node {$a\mid g$}()
    (q1) edge [loop above]  node {$a \mid g$}()
    (qn) edge [loop above]  node {$a \mid g$}();

\end{tikzpicture}
        \caption{The Mealy automaton $\A$ that embeds $\Z^2$ into $(G,S\cup\{g\})$. The notation $s\mid t$ represents $\eta(q,s) = t$ for the corresponding state.}
        \label{fig:snake_embedding_lemma}
    \end{figure}
    
    Let $f = f_{\A}$ be the function associated to $\A$. Take $v\in\{a,b,a^{-1},b^{-1}\}^{*}$. As $\Z^2$ is abelian, we can express $v$ in the normal form $v =_{\Z^2} a^kb^{l}$ with $k = |v|_a - |v|_{a^{-1}}$ and $l = |v|_{b} - |v|_{b^{-1}}$. If we prove that $f_{\A}(v) =_G f_{\A}(a^kb^l)$, then the function $\phi(g) = f_{\A}(v)$, for any $v$ representing $g\in\Z^2$, defines a snake-embedding. Indeed, if we have such identity, because any two words $v_1$ and $v_2$ are equal in $\Z^2$ if and only if they have the same normal form, $\phi$ will be well-defined and injective.

    Let $x = w^{\infty}$ be the periodic configuration that will specify one of copies of $\Z$ in $G$. By the transducer's definition, $f_{\A}(a^kb^l) = g^{k}x_{[0,l-1]}$. We will show that $f_{\A}(v) =_{G} g^{k}x_{[0,l-1]}$ through induction on the length of $v$. If $v = \epsilon$, then $f_{\A}(\epsilon) =_G 1_{G}$. Next, suppose the equality is true for all words $u$ such that $|u|\leq n$. We arrive at different cases:
    \begin{itemize}
        \item If $v = v'a$ and $v'$ has normal form $a^kb^{l}$, then $f_{\A}(v) = f_{\A}(v')\eta(q_{l\text{ mod }m}, a) = f_{\A}(v')g$. By induction, we know $f_{\A}(v') = g^{k}x_{[0,l-1]}$ and thus $f_{\A}(v) = g^{k}x_{[0,l-1]}g$. But, as $g$ is in the center of $G$, we can make it commute with $x_{[0,l-1]}$ arriving at $f_{\A}(v) =_{G} g^{k+1}x_{[0,l-1]}$.

        \item If $v = v'b$ and $v'$ has normal form $a^kb^{l}$, then $f_{\A}(v) = f_{\A}(v')\eta(q_{l\text{ mod }m}, b) = f_{\A}(v')w_{l\text{ mod }m}$. By induction, we know $f_{\A}(v') = g^{k}x_{[0,l-1]}$ and thus $f_{\A}(v) = g^{k}x_{[0,l-1]}w_{l\text{ mod }m}$. But by definition $x_{[0,l]} = x_{[0,l-1]}w_{l\text{ mod } m}$, therefore $f_{\A}(v) =_{G} g^{k+1}x_{[0,l]}$.
    \end{itemize}
\end{proof}

\begin{mylem}{\ref{lem: mso_ray}}
Let $P\subseteq V$. Then, 
\begin{enumerate}
    \item If there exists a partition $\{P_{s,a}\}_{s\in S, a\in A}$ such that $\infty\textsc{ray}(P,\{P_{s,a}\})$ is satisfied, $P$ contains an infinite injective path. Conversely, if $P$ is the support of an injective infinite path rooted at $v_0$, there exists a partition $\{P_{s,a}\}_{s\in S, a\in A}$ such that $\infty\textsc{ray}(P,\{P_{s,a}\})$ is satisfied.
    \item If there exists a partition $\{P_{s,a}\}_{s\in S, a\in A}$ such that $\textsc{loop}(P,\{P_{s,a}\})$ is satisfied, $P$ contains a simple loop. Conversely, if $P$ is the support of a simple loop based at $v_0$, there exists a partition $\{P_{s,a}\}_{s\in S, a\in A}$ such that $\textsc{loop}(P, \{P_{s,a}\})$ is satisfied. 
\end{enumerate}
\end{mylem}

\begin{proof}
\begin{enumerate}
    \item Suppose $\infty\textsc{ray}(P, \{P_{s,a}\})$ is satisfied, for some partition $\{P_{s,a}\}$. We will recursively define an injective 1-Lipschitz function $f:\N\to P$ that will give us our path. Start by setting $f(0) = v_0$. Because the formula is satisfied, there exists a unique $s\in S$ such that $v_0\cdot s \in P$ and we define $f(1) = v_0\cdot s$. Now, suppose we have already defined $f(n)$. Then, there exists a unique $s'\in S$ such that $f(n)\cdot s'\in P$, so we set $f(n+1) = f(n)\cdot s'$. Thus, $f$ is well-defined. In addition, $f$ is injective because if there were $n,m\in \N$ such that $v = f(n) = f(m)$, $v$ would have two distinct predecessors. 

    Next, if $P$ supports an infinite injective path given by $f:\N\to P$ with $f(0) = v_0$, for all $n\in \N$ there exists $s_n\in S$ such that $f(n+1) = f(n)\cdot s_n$. We define the sets $P_{s,a} = \{f(n)\mid s_n = s\}$ for a fixed $a\in A$, which partition $P$. Finally, as $f$ is injective, every $v\in P$ has a unique predecessor and therefore $\infty\textsc{ray}(P, \{P_{s,a}\})$ is satisfied.

    \item Suppose $\textsc{loop}(P, \{P_{s,a}\})$ is satisfied, for some partition $\{P_{s,a}\}$. Then, as we did for the infinite path case, we can define the function $l:\llbracket 0, n\rrbracket \to P$ with $l(0) = l(n) = v_0$, for some $n\geq 3$ by using the fact that $P$ satisfies $N(P, \{P_{s,a}\})$ (starting from $v_0$ we can always find a successor) and that for every subset $Q\subseteq P$ the formula $\forall\{Q_{s,a}\}\neg\infty\textsc{ray}(Q,\{Q_{s,a}\})$ is satisfied (which tells us that $P$ cannot contain the support of an infinite ray) we know such an $n$ must exist. As before we know $l$ defines a simple loop because $P$ satisfies $\ell(P, \{P_{s,a}\})$.

    Finally, let $P$ supports a simple loop defined by $l:\llbracket0,n\rrbracket\to P$ with $l(0) = l(n) = v_0$. By definition, for every $i\in \llbracket0,n-1\rrbracket$, there exists $s_i\in S$ such that $l(i+1) = l(i)\cdot s_i$. Thus, the partition defined by the sets $P_{s,a} = \{\ell(i) \mid s_i = s\}$ for a fixed $a\in A$ satisfies the required properties in virtue of $P$ being a simple non-trivial loop.
\end{enumerate}
\end{proof}

\section*{Appendix B: Equivalence between Wang tiles and Tileset graphs}

\begin{definition}
\label{wang}
Let $G$ be a f.g. group, $S$ a finite set of generators, $C$ a finite set of colors. A \emph{Wang tile} for $(G,S)$ is an element $\theta\in C^{S\cup S^{-1}}$.
\end{definition}

Let $(G,S)$ be a f.g. group and $\tau$ a Wang tile set for the group. A $\tau$-snake is a pair of functions $(\om, \zeta)$ such that $\om:I\to G$ is a skeleton, $\zeta:I\to \tau$ and $\zeta(i)_{d\om_i} =\zeta(i+1)_{d\om_{i}^{-1}}$.

\begin{lemma}
Let $\tau$ be a Wang tileset. and $(\om, \zeta)$ a $\tau$-snake. Then, there exists a tileset graph $\Gamma_{\tau}$, effectively constructed from $\tau$, such that $(\om, \zeta)$ is a $\Gamma_{\tau}$-snake if and only if $(\om, \zeta)$ is a $\tau$-snake.
\end{lemma}

\begin{proof}
 Given a Wang tile set, we can directly construct a finite graph $\Gamma_{\tau}$ with vertex set $\tau$ such that there is an edge labeled $s\in(S\cup S^{-1})$ from $\theta_1$ to $\theta_2$ if and only if $(\theta_{1})_{s} = (\theta_{2})_{s^{-1}}$. This makes $\Gamma_{\tau}$ a tileset graph. Then, $(\om, \zeta)$ is a $\Gamma_{\tau}$-snake if and only if it is a $\tau$-snake as $\Gamma_{\tau}$ was created to preserve the adjacency rules of the Wang tiles.
\end{proof}

\begin{lemma}
Let $\Gamma$ be a tileset graph. Then, there exists a Wang tile set $\tau_{\Gamma}$, effectively constructed from $\Gamma$, such that $(\om, \zeta)$ is a $\Gamma$-snake if and only if there exists $\zeta_{\Gamma}:I \to\tau_{\Gamma}$ such that $(\om, \zeta_{\Gamma})$ is a $\tau_{\Gamma}$-snake.
\end{lemma}

\begin{proof}
 Now, let $\Gamma = (A, B)$ be a nearest neighbor graph. Given $a\in A$, we say that $N\in ( A\cup \{\square\})^{S\cup S^{-1}}$ is a neighborhood of $a$ if for all $s\in S\cup S^{-1}$ such that there is an outgoing edge labeled $s$ from $a$, $(a, N_{s}, s)$ is an edge in $\Gamma$, and $N_s = \square$ when there is no such edge. Next, we define the set of colors $C$ for our new Wang tiles as
 $$C = \{(a,b,s) \mid (a,b,s) \text{ is an edge on }\Gamma\}\cup D,$$
 where $D$ is a finite set of colors representing tiles where edges of a particular generator are not present. Finally, $\tau_{\Gamma}$ is composed of tiles $\theta_{a, N}$, for each $a\in A$ and $M$ a neighborhood of $a$. Edges are colored according to $N$ as $(\theta_{a,N})_s = (a,N_s,s)$ if $s\in S$ and $N_s\neq \square$, $(\theta_{a,N})_{s^{-1}} = (N_{s^{-1}}, a,s)$ if $s\in S$ and $N_{s^{-1}}\neq \square$ and $(\theta_{a,N})_{s^{\pm1}}\in D$ otherwise. This tileset can be computed from a description of $\Gamma$, as it suffices to enumerate all possible neighborhoods for each tile and construct the corresponding Wang tiles.
 
 Now, let $(\om, \zeta)$ be a $\Gamma$-snake. Define $\zeta_\Gamma(i) = \theta_{\zeta(i), N_i}$ with $N_i$ being any neighborhood of $\zeta(i)$ such that $(N_i)_{d\om_{i-1}^{-1}} = \zeta(i-1)$ and $(N_i)_{d\om_i} = \zeta(i)$. Thus, $(\om, \zeta_\Gamma)$ is a $\tau_{\Gamma}$-snake. Conversely, if there is a map $\zeta_{\Gamma}:I\to\tau_{\Gamma}$ such that $(\om, \zeta_{\Gamma})$ is a $\tau_{\Gamma}$-snake, for every $i\in I$ there is a tile $a_i\in A$ and a neighborhood $N_i$ of $a_i$ such that $\zeta_{\Gamma}(i) = \theta_{a_i, N_i}$. Define $\zeta(i) = a_i$, Then, for any $i\in I$, if we suppose without loss of generality that $d\om_i\in S$, we have that 
 \begin{align*}
     (a_i, (N_i)_{d\om_i}, d\om_i) &= \theta_{a_i, N_i}(d\om_i),\\
     &=\theta_{a_{i+1}, N_{i+1}}(d\om_i^{-1}),\\
     &= ((N_{i+1})_{d\om_i^{-1}}, a_{i+1}, d\om_i),
 \end{align*}
 As $\zeta_{\Gamma}(i)$ and $\zeta_{\Gamma}(i+1)$ are Wang tiles adjacent through $d\om_i$. Thus $(N_i)_{d\om_i} = a_{i+1}$, which implies $(\zeta(i), \zeta(i+1))$ is an edge labeled $d\om_i$ in $\Gamma$, making $(\om, \zeta)$ a $\Gamma$-snake.

\end{proof}

\end{document}